\documentclass[journal]{IEEEtran}

\makeatletter
\newcommand*\titleheader[1]{\gdef\@titleheader{#1}}
\AtBeginDocument{%
  \let\st@red@title\@title
  \def\@title{%
    \bgroup\normalfont\large\centering\@titleheader\par\egroup
    \vskip0.0em\st@red@title}
}
\makeatother

\usepackage{blindtext,graphicx,amsmath,bm,subcaption,multirow,tabularx,caption,ragged2e,booktabs, verbatim, tabu,cite, amsthm}
\usepackage[utf8]{inputenc}
\usepackage{enumitem}
\usepackage{array}
\usepackage{makecell}
\usepackage{indentfirst,mathtools}
\usepackage{bbold,setspace}
\usepackage[font=footnotesize]{caption}
\usepackage{xcolor,color}
\usepackage{MnSymbol}
\usepackage{wasysym}
\usepackage[normalem]{ulem}
\newcommand\redsout{\bgroup\markoverwith{\textcolor{red}{\rule[0.5ex]{2pt}{1pt}}}\ULon}
\usepackage[utf8]{inputenc}
\usepackage{enumitem}
\usepackage{mathtools}

\newcommand{\comm}[1]{}
\ifCLASSINFOpdf
 
\else
 
\fi

\begin{document}

\title{Impact of Virtual Inertia on DC Grid Stability with Constant Power Loads}

\author{Hao~Tu,~\IEEEmembership{Member,~IEEE,}
Hui~Yu,~\IEEEmembership{Member,~IEEE,}
     and~Srdjan~Lukic,~\IEEEmembership{Senior~Member,~IEEE\vspace{-20pt}}
\thanks{Hao Tu, Hui Yu, and Srdjan Lukic are with North Carolina State University, Raleigh, NC 27695, USA (email: htu@ncsu.edu; hyu11@ncsu.edu; smlukic@ncsu.edu).
}
}

\maketitle
\begin{abstract}

Virtual inertia is an effective control approach to attenuate sudden voltage changes during transient events in low-inertia DC grids. While methods have been proposed to implement virtual inertia, its impact on DC grid stability in the presence of constant power loads (CPLs) remains unclear. In this paper, we perform a rigorous stability analysis for DC grids with CPLs powered by virtual-inertia-enhanced converters. We derive a closed-form stability criterion that can be used to evaluate the impact of virtual inertia on the system stability, and demonstrate that, given a set of system parameters, the stability of a DC grid powering CPLs can be improved for a range of virtual inertia designs. We provide analytical expressions for the optimal virtual inertia that improves stability and for the maximum virtual inertia that does not deteriorate stability. In addition, we present a step-by-step guideline to design a stable DC grid with virtual inertia. Test results are presented to validate the analysis.  
\end{abstract}

% Note that keywords are not normally used for peerreview papers.
\IEEEpeerreviewmaketitle
\vspace{-10pt}
\section{Introduction}

\IEEEPARstart{I}{n} DC grids, droop control is commonly used to enable multiple source converters to share load without communication~\cite{hierarchical}. Droop control is a static equation without any internal states: any change in the load will be directly reflected on the terminal voltage, which leads to rapid voltage fluctuations during abrupt load changes. Analogous to controlling the frequency fluctuation in AC grids, virtual inertia can attenuate voltage magnitude fluctuations~\cite{7015592, 7801819,samanta2018virtual,9069316, 9264685} in DC grids. In \cite{7015592}, a first-order low-pass filter~(LPF) is added to droop control to provide virtual inertia in DC grids. In~\cite{7801819,samanta2018virtual,9069316, 9264685}, virtual inertia is implemented in DC grids by emulating the swing equation of an electric machine.

In DC grids, the negative incremental impedance of constant power loads~(CPLs) may cause instability. In \cite{kwasinski2010dynamic}, DC grid stability is analyzed for a buck converter powering a CPL. Several methods are discussed to enhance stability such as shedding CPLs, increasing DC bus capacitance, or employing advanced passivity-based control. In \cite{9707664}, DC system stability is analyzed using an impedance-based criterion, and a control method is proposed to shape the impedance of the CPL. In \cite{5764841}, additional passive components are added to improve the system stability. In \cite{6909049}, the CPL stability problem is analyzed for DC grids formed by droop-controlled converters. The authors conclude that a system equilibrium can become unstable if the DC bus capacitance is smaller than a threshold. In~\cite{6948369}, the droop gain is interpreted as a virtual resistor, and its impact on stability is discussed. In~\cite{6031929}, tools for large-signal stability analysis are presented to determine the region of attraction (ROA) of equilibrium points. In \cite{9855660}, large-signal stability analysis is performed using Brayton-Moser's mixed potential theory. A method is proposed to increase the ROA in \cite{7182770}.

An open question is how virtual inertia affects the stability of a DC grid dominated by CPLs. In~\cite{7015592, 7801819,samanta2018virtual,9069316, 9264685}, stability with virtual inertia is evaluated numerically, demonstrating that the system under study is stable. In \cite{6031929,9707664,kwasinski2010dynamic,5764841,6948369,6909049,9855660,7182770}, virtual inertia is not considered in the stability analysis. In \cite{7539344}, DC grid stability with CPLs is analyzed with a frequency-dependent virtual impedance, which is equivalent to the LPF implementation of virtual inertia. The effect of varying virtual inertia on stability is evaluated numerically for the system under study. Results in \cite{7539344} show that the stability of the studied system can be improved by virtual inertia. However, it is unclear whether this conclusion can be generalized to other DC grids designs, and no closed-form design metric is provided to guide practicing engineers in selecting the appropriate virtual inertia. 

In this paper, we perform a rigorous stability analysis and show how virtual inertia impacts DC grid stability with CPLs. First, we show that the LPF and machine emulation implementations of virtual inertia in DC grids are equivalent. Next, we formulate the criterion for evaluating DC grid stability when virtual inertia is considered. We prove that the stability criterion can be reduced to a single closed-form expression. We evaluate the impact of virtual inertia on stability with CPLs. Specifically, we derive the analytical expressions for the optimal virtual inertia that improves stability and for the maximum virtual inertia that can be provided without deteriorating stability.  Finally, we present a step-by-step guideline to design a stable DC grid with virtual inertia.

The rest of the paper is organized as follows. In Section~\ref{problem}, a representative DC system for stability analysis is introduced, and different implementations of virtual inertia are discussed. Section~\ref{stability} presents the stability analysis. In Section~\ref{design_sec}, the design guideline is presented. Section~\ref{experiment} shows the test results and Section \ref{conclusion} concludes the paper.

\section{Representative DC System with Virtual Inertia} \label{problem}

\begin{figure}[t]
  \centering
	\includegraphics[width=0.33\textwidth]{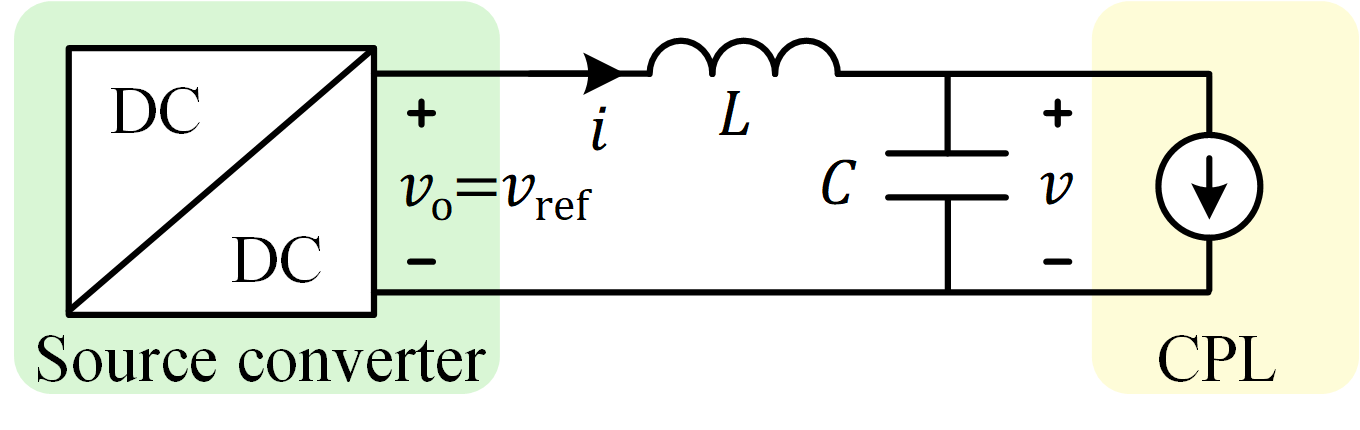}
	\vspace{0pt}
	\caption{Representative system for stability analysis of DC grids.}
	\label{fig:system}
	\vspace{-5pt}
\end{figure}

We adopt a representative system widely used in literature~(e.g. \cite{6031929,9707664,kwasinski2010dynamic,5764841,6909049,7182770}) to study DC grid stability with CPLs~(see Fig.~\ref{fig:system}). The system consists of one source converter with virtual inertia control and one CPL, connected through an inductor and a capacitor. The source converter can be interpreted as an aggregated model of multiple paralleled converters using the method in~\cite{6909049}. The inductor and capacitor can be seen as the equivalent line impedance, filters, and DC bus capacitor. To eliminate the stabilizing effects of resistive components~\cite{5764841}, we ignore any parasitic resistance and assume that there is no resistive load on the bus. We assume that the converter output voltage is accurately regulated to the reference voltage by the inner control loops, i.e., $v_o=v_\mathrm{ref}$, for the theoretical analysis.

\begin{figure}
     \centering
     \begin{subfigure}[b]{0.19\textwidth}
         \centering
         \includegraphics[width=\textwidth]{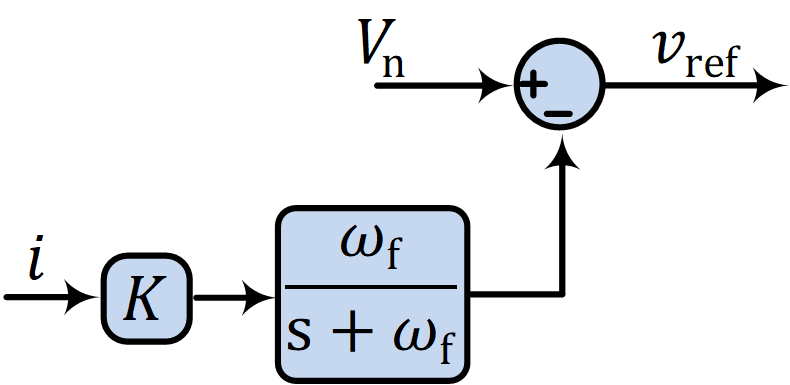}
         \caption{Low-pass filter approach.}
         \label{fig:lpf}
     \end{subfigure}
     \begin{subfigure}[b]{0.26\textwidth}
         \centering
         \includegraphics[width=\textwidth]{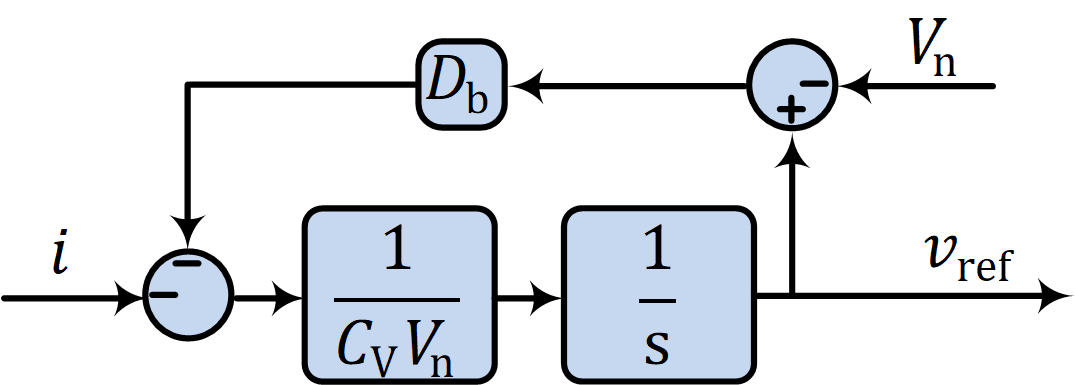}
         \caption{Machine emulation approach.}
         \label{fig:machine}
     \end{subfigure}
        \caption{Different implementations of virtual inertia in DC grids.}
        \label{fig:inertia}
        \vspace{-10pt}
\end{figure}

For DC grids, $v_\mathrm{ref}$ is given by droop control as~\cite{6531679},
\begin{equation}\label{droop}
     v_\mathrm{ref}= V_\mathrm{n} - K i,
\end{equation}   
\noindent where $K$ is the droop gain, $V_\mathrm{n}$ is the nominal voltage of the DC grid, and $i$ is the source converter's output current. Droop control~\eqref{droop} for DC grids is a static equation without internal dynamics. Any fluctuation in $i$ will directly change $v_\mathrm{ref}$. To prevent rapid voltage fluctuations, the concept of virtual inertia is introduced. Analogous to the method of implementing virtual inertia in AC grids, in \cite{7015592}, a first-order LPF is added to provide virtual inertia in DC grids as shown in Fig.~\ref{fig:inertia}(a). The dynamics are described by
\begin{equation}\label{inertia1}
    \dot  v_\mathrm{ref}=\omega_\mathrm{f}( V_\mathrm{n} -v_\mathrm{ref}) -\omega_\mathrm{f} K i,
\end{equation} 
where $\omega_\mathrm{f}$ is the LPF bandwidth.

In~\cite{7801819,samanta2018virtual,9069316, 9264685}, the authors implement virtual inertia in DC grids by emulating a DC equivalent synchronous machine. Fig.~\ref{fig:inertia}(b) shows a typical machine emulation method given by~\cite{7801819}. The dynamics are described by
\begin{equation}\label{inertia2}
     C_\mathrm{v}V_\mathrm{n} \dot v_\mathrm{ref}= -D_\mathrm{b}(v_\mathrm{ref} -V_\mathrm{n}) - i,
\end{equation} 
where $C_\mathrm{v}$ is the emulated virtual inertia and $D_\mathrm{b}$ is the emulated damping factor. It can be observed that \eqref{inertia1} and \eqref{inertia2} are equivalent by selecting $\omega_\mathrm{f} = D_\mathrm{b}/(C_\mathrm{v}V_\mathrm{n})$ and $K=1/D_\mathrm{b}$. In this paper, we will use \eqref{inertia1} for the analysis but the conclusions are valid for both implementations of virtual inertia.

When the source converter is enhanced with virtual inertia, the dynamics of the system in Fig.~\ref{fig:system} are described by
\begin{subequations}\label{system_model}
\begin{align}
    \label{system_model1}
    \dot  v_\mathrm{ref} &=\omega_\mathrm{f}( V_\mathrm{n} -v_\mathrm{ref}) -\omega_\mathrm{f} K i, \\
    \label{system_model2}
    L\dot i &= v_\mathrm{ref} - v, \\ 
    \label{system_model3}
    C \dot v & = i - P/v,
\end{align}
\end{subequations}
where $P$ is the CPL power and $v$ is the capacitor voltage. 

In~\eqref{system_model}, we model the source converter as an ideal voltage source and use an ideal CPL model. This approach is equivalent to assuming that the converter bandwidths are infinite, or in other words, that the dynamics of the converters do not influence the droop and virtual inertia dynamics. This approximation is well accepted in the literature as the ideal models facilitate the theoretical analysis while capturing the critical dynamics that cause stability issues~\cite{6031929}.

\section{Stability Analysis }\label{stability}

\subsection{Stability Criterion}
To determine the system equilibrium points, we let the derivatives in \eqref{system_model} be zero.  Then, \eqref{system_model1} becomes
\begin{align}
    0=V_\mathrm{n} -v_\mathrm{ref} - K i,
\end{align}
which is the same as droop control \eqref{droop} without virtual inertia. This shows that virtual inertia does not change the system equilibrium points. The system equilibrium points under droop control have been discussed in~\cite{6909049}. Here, we give the steady-state reference voltage $V_\mathrm{ref,e}$, capacitor voltage $V_\mathrm{e}$, and inductor current $I_\mathrm{e}$ directly,
\begin{align} \label{Ve}
    V_\mathrm{ref,e} = V_\mathrm{e}&=  \frac{V_\mathrm{n}+\sqrt{V_\mathrm{n}^2-4PK}}{2}, \ \ 
    I_\mathrm{e} = \frac{V_\mathrm{n}-V_\mathrm{e}}{K}.
\end{align}
Two important facts can be observed from~\eqref{Ve}. First, $V_\mathrm{e}$ is a monotonically decreasing function of $P$. Second, the equilibrium point given by~\eqref{Ve} exists if
\begin{align}\label{Pmax}
P\leq P_\mathrm{max}=V_\mathrm{n}^2/(4K),
\end{align}
where $P_\mathrm{max}$ is commonly known as the power transfer limit via a (virtual) resistor of resistance $K$, since the droop gain can be interpreted as a virtual resistor in series with $V_\mathrm{n}$.

The local stability of the equilibrium point is determined by  the Jacobian matrix at the equilibrium point, which is
\begin{align} \label{J}
J = \begin{bmatrix} 
    -\omega_\mathrm{f} & -\omega_\mathrm{f} K & 0 \\
     1/L & 0 & -1/L \\
     0 & 1/C & 1/(R_\mathrm{e}C)\
     \end{bmatrix} \ \ \mathrm{with} \ \ R_\mathrm{e} = \frac{V_\mathrm{e}^2}{P}>0.
\end{align}
Commonly, $-R_\mathrm{e}$ is referred to as the incremental (or small-signal) impedance of the CPL, and it behaves like a negative resistor to destabilize the system. Because $R_\mathrm{e}$ is a monotonically decreasing function of $P$, equation \eqref{Pmax} provides a lower bound for $R_\mathrm{e}$,
\begin{align} \label{Re}
R_\mathrm{e} \geq K.
\end{align}

The eigenvalues of $J$ are the roots of the following equation, 
\begin{align}
\lambda^3+a_2\lambda^2 + a_1\lambda + a_0 = 0,
\end{align}
where,
\begin{align}
    \nonumber
    a_2 &= \omega_\mathrm{f} - \frac{1}{R_\mathrm{e}C}, \ \  \
    a_1 =  \frac{R_\mathrm{e} + \omega_\mathrm{f} K C R_\mathrm{e} -\omega_\mathrm{f} L}{LCR_\mathrm{e}},\\\label{factors}
    a_0 &=  \frac{\omega_\mathrm{f} R_\mathrm{e}- \omega_\mathrm{f} K}{LCR_\mathrm{e}}.
\end{align}
According to Routh–Hurwitz criterion, all the eigenvalues have negative real parts if and only if
\begin{align} \label{Routh}
    a_2 > 0, \ \ \
    a_1 > 0,\ \ \  \mathrm{and}  \ \ \
    a_2a_1-a_0 >0.
\end{align}
Substituting \eqref{factors} into \eqref{Routh} gives the stability criterion as
\begin{subequations}\label{factor_Routh}
\begin{align} 
    f_2 &=  \omega_\mathrm{f} R_\mathrm{e}C -1 >0, \\
    f_1 &=  R_\mathrm{e} + \omega_\mathrm{f} K C R_\mathrm{e} -\omega_\mathrm{f} L
    >0,\\
    f_0 &=  \omega_\mathrm{f}^2 K C^2 R_\mathrm{e}^2 - \omega_\mathrm{f}^2 L C R_\mathrm{e} + \omega_\mathrm{f} L -R_\mathrm{e} >0.
\end{align}
\end{subequations}
For a specific system, if all the variables are known, \eqref{factor_Routh} can be used to evaluate the system stability.

\vspace{-5pt}
\subsection{Proposed Stability Criterion}
While \eqref{factor_Routh} is a rigorous stability criterion, it provides little insight into the virtual inertia's impact on stability. To provide deeper insight, we explore the effect of $C$ and $\omega_\mathrm{f}$ on stability, while keeping $K$, $R_\mathrm{e}$, and $L$ constant. As discussed in \cite{kwasinski2010dynamic} and \cite{6909049}, one of the most common practices to improve stability in the presence of CPLs is to increase $C$. This is because all the other parameters are difficult to vary significantly: droop gain $K$ is determined by the desired steady-steady voltage deviation~\cite{6953474}; $R_\mathrm{e}$ is determined by $K$, $P$, and $V_\mathrm{n}$, which are fixed by the grid operating condition; $L$ is difficult to change in a wide range. It can also be observed that the criterion \eqref{factor_Routh} is always satisfied if $C$ is large enough. Based on the discussion above, we convert \eqref{factor_Routh} into requirements for $C$ as
\begin{subequations}\label{C_Routh}
\begin{align} \label{C_Routh1}
    r_2 &:= \{C \mid C > C_2, \  \mathrm{with} \ C_2=1/(\omega_\mathrm{f} R_\mathrm{e})\}, \\\label{C_Routh2}
    r_1 &:=  \{C \mid C >C_1, \  \mathrm{with} \ C_1=L/(KR_\mathrm{e})-1/(\omega_\mathrm{f} K)\} \\\label{C_Routh3}
    r_0 &:=\{C \mid  C < C_{0}^- \ \ \mathrm{or}  \ \ C > C_0 \},
\end{align}
\end{subequations}
where $C_{0}^-$ and $C_{0}$ are the roots for the equation $f_0(C) = 0$,
\begin{align} \label{C0-}
     C_{0}^- & = \frac{L-\sqrt{L^2-4K(\frac{L}{\omega_\mathrm{f}}-\frac{R_\mathrm{e}}{\omega_\mathrm{f}^2})}}{2KR_\mathrm{e}},\\ \label{C0}
     C_0 & = \frac{L+\sqrt{L^2-4K(\frac{L}{\omega_\mathrm{f}}-\frac{R_\mathrm{e}}{\omega_\mathrm{f}^2})}}{2KR_\mathrm{e}}.
\end{align}
With~\eqref{Re}, the term under the square root in \eqref{C0-} and \eqref{C0} is always positive. Therefore, $C_{0}^-$ and $C_{0}$ always exist. Criterion \eqref{factor_Routh} is satisfied if $C \in r_2\cap r_1 \cap r_0$, which can be reduced to a single requirement of $C$ as we show in the following criterion. 

\vspace{+3pt}
\textbf{Proposed Stability Criterion.}  \textit{The equilibrium point given by \eqref{Ve} of the system~\eqref{system_model} is }
\begin{enumerate}
    \item \textit{locally stable if $C \in r_s , \  where \  r_s =\{C \mid C > C_0 \}$};

\item \textit{unstable if $C \in r_u , \  where \  r_u =\{C \mid C < C_0 \}.$}
\end{enumerate}

\begin{proof} 1) To prove the first part of the proposed stability criterion, we show $ r_s =r_2\cap r_1 \cap r_0$. We begin with $r_2\cap r_0=r_s$. Substituting $C_2$ into $f_0(C)$ gives
\begin{align} \label{fc2}
    f_0(C_2) = K -R_\mathrm{e} \leq0,
\end{align}
where \eqref{Re} is used. Since $C_{0}^-$ and $C_{0}$ are the roots for the quadratic equation $f_0(C) = 0$, \eqref{fc2} means
\begin{align} \label{c0c2c0}
    C_{0}^- \leq C_2 \leq C_0
\end{align}
for any $\omega_\mathrm{f} \in (0, +\infty)$. The intersection of $r_2$ and $r_0$ becomes
\begin{align} \label{C_Routh4}
    r_2\cap r_0 =\{C \mid C > C_0 \}=r_s.
\end{align}

Next, we show $r_s\cap r_1=r_s$ by showing $ C_0 \geq C_1$ for any $\omega_\mathrm{f} \in (0, +\infty)$. We let $\tau = 1/\omega_\mathrm{f}$ and
\begin{align} \nonumber
    g(\tau) &= C_0(\tau)-C_1(\tau) \\ \label{g}  &=\frac{L+\sqrt{L^2-4K(L\tau-R_\mathrm{e}\tau^2)}}{2KR_\mathrm{e}}+\frac{\tau}{K}- \frac{L}{KR_\mathrm{e}}.
\end{align}
The derivative of $g(\tau)$ with respect to $\tau$ is
\begin{align} \label{g'}
    g'(\tau) = \frac{2\tau R_\mathrm{e}-L}{R_\mathrm{e}\sqrt{L^2-4K(L\tau-R_\mathrm{e}\tau^2)}} +\frac{1}{K}.
\end{align}

We allow $\tau \in [0, +\infty)$ for now. By observing $g(\tau=0)=0$~(i.e. $C_0 = C_1$ for $\tau=0$), to obtain $C_0 \geq C_1$ for any $\tau \in [0, +\infty)$, it is sufficient to show that $g(\tau)$ is a non-decreasing function of $\tau$, i.e. $g'(\tau) \geq0$ for any $\tau \in [0, +\infty)$. 

We show that $g'(\tau)\geq0$ for $\tau \in [0, \frac{L}{2R_\mathrm{e}})$ and $\tau \in [\frac{L}{2R_\mathrm{e}}, +\infty)$ separately. If $\tau \in [\frac{L}{2R_\mathrm{e}}, +\infty)$, $g'(\tau)>0$ since the first term of \eqref{g'} is non-negative and the second term is positive. 

If $\tau \in [0, \frac{L}{2R_\mathrm{e}})$, by applying a change of variable 
\begin{align}
y = \frac{L}{2R_\mathrm{e}} - \tau \ \ \mathrm{with} \ \  y \in (0, \frac{L}{2R_\mathrm{e}}], 
\end{align}
equation \eqref{g'} becomes 
\begin{align} \label{g'y}
    g'(y) = \frac{-2}{\sqrt{4KR_\mathrm{e}+\frac{L^2(R_\mathrm{e}-K)}{R_\mathrm{e}y^2}}} +\frac{1}{K}.
\end{align}
With~\eqref{Re}, \eqref{g'y} shows that $g'(y)$ is a monotonically decreasing function of $y$ for $y \in (0, \frac{L}{2R_\mathrm{e}}]$, which means
\begin{align}
    g'(y)\geq g'(y=\frac{L}{2R_\mathrm{e}})=\frac{1}{K}-\frac{1}{R_\mathrm{e}}\geq0
\end{align}
in this range, and the equality holds only at $y=\frac{L}{2R_\mathrm{e}}$~(i.e. $\tau=0$) when $K=R_\mathrm{e}$.

Combining the above results, it is concluded that $g'(\tau) \geq0$ for any $\tau \in [0, +\infty)$ with that equality holds only at $\tau=0$ when $K=R_\mathrm{e}$. Therefore, $g(\tau)\geq g(0) =0$ for any $\tau \in [0, +\infty)$ with equality holds only at $\tau=0$. Considering $g(\tau) = C_0(\tau)-C_1(\tau)$ and $\tau = 1/\omega_\mathrm{f}$, this means $C_0\geq C_1$ for any  $\omega_\mathrm{f} \in (0, +\infty)$. The proof for the first part of the criterion is completed.

2) To prove the second part, it is sufficient to show that at least one requirement of \eqref{C_Routh} is violated if $C \in r_u$. Since \eqref{c0c2c0} holds, the following two cases are discussed.

Case 1~($C_{0}^- \leq C_2 < C_0$): In this case, requirement~\eqref{C_Routh1} is violated if $C\in [0, C_2)$; requirement~\eqref{C_Routh3} is violated if $C\in [C_2, C_0)$.  Therefore, \eqref{C_Routh1} or \eqref{C_Routh3} is violated if $C \in r_u$.

Case 2~($C_{0}^- \leq C_2 = C_0$): In this case, requirement~\eqref{C_Routh1} is always violated if  $C \in r_u$.\end{proof}

The proposed stability criterion shows that $C_0$ is the capacitance at the boundary between stability and instability. In practice, it is difficult to operate the system near the boundary because the equilibrium near the boundary has a small ROA~\cite{6031929,7182770}, meaning that a small perturbation can make the system unstable. If \eqref{C0} is used to design the capacitance, some margin is necessary for practical implementations. The design procedure is discussed in Section~\ref{design_sec} in detail.

\vspace{-5pt}
\subsection{Impact of Virtual Inertia}

\begin{figure}[t]
  \centering
	\includegraphics[width=0.42\textwidth]{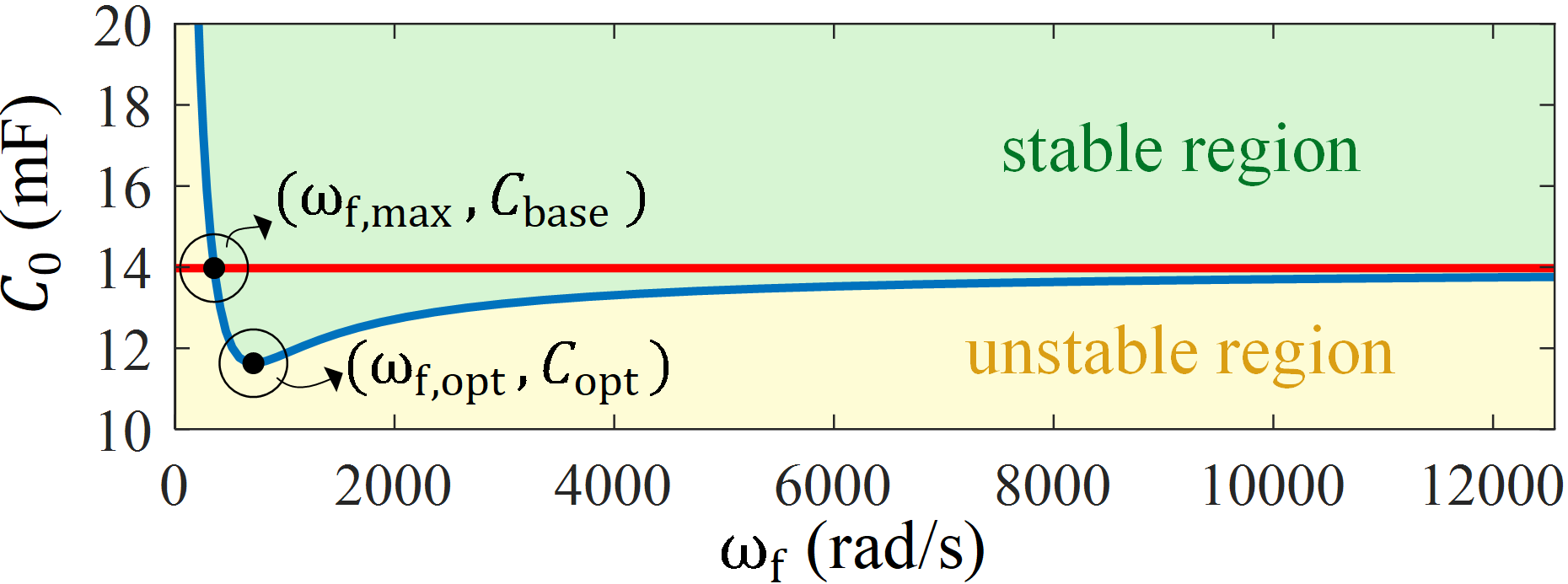}
	\vspace{-3pt}
	\caption{Blue - $C_0$ as a function of the LPF bandwidth $\omega_\mathrm{f}$, generated using parameters given in Table~\ref{Table1}; red - $C_{\mathrm{base}}$ without virtual inertia. }
	\label{fig:C0}
	\vspace{-4pt}
\end{figure}

With the stability criterion reduced to a single constraint, we can evaluate the impact of virtual inertia on the required capacitance~\eqref{C0} to guarantee stability. We plot the required capacitance $C_0$ as a function of the LPF bandwidth $\omega_\mathrm{f}$ in Fig.~\ref{fig:C0} and discuss some of the salient points below.

\textbf{Baseline}: We first evaluate the required capacitance for stability as  $\omega_\mathrm{f}\to\infty$,
\begin{align} \label{cbaseline}
    C>C_{\mathrm{base}} = C_0(\omega_\mathrm{f}\to\infty) = L/(KR_\mathrm{e}).
\end{align}
With an LPF of infinite bandwidth (i.e. a system with no LPF), the system has no inertia and the dynamics 
converge to those under droop control. The same requirement as \eqref{cbaseline} has been identified for droop control without virtual inertia in the literature (e.g. \cite{6909049, 7182770,kwasinski2010dynamic}).

\begin{table}[t]
\footnotesize
\centering\captionsetup{justification=centering}
\caption{{System parameters for single-converter case}}\label{Table1}
\setlength\tabcolsep{4pt} % default value: 6pt
\begin{tabular}{ccccc|cc}%{|p{2.5cm}|p{2.5cm}|p{2.5cm}|}
\hline
\hline
\multicolumn{5}{c|}{Grid parameters}  &\multicolumn{2}{c}{Virtual inertia $(\mathrm{rad/s})$}
\\
\hline
$V_\mathrm{n} (\mathrm{V})$ & $K$ & $L (\mathrm{mH})$ & $C (\mathrm{mF})$ & $P (\mathrm{kW})$ & \ $\omega_\mathrm{f,opt}$ &$\omega_\mathrm{f,max}$\\
\hline
$200$ &$0.2$ & $1$ & $14$ & $46$ & $715$& $357$\\
\hline
\hline
\end{tabular}
\vspace{-13pt}
\end{table}

\textbf{Optimal Inertia (for stability)}: By reducing LPF bandwidth $\omega_\mathrm{f}$, virtual inertia is introduced into the system. Lower LPF bandwidth means larger virtual inertia, and vice versa. Equation \eqref{C0} shows that $C_0$ is a decreasing function for $\omega_\mathrm{f} \in (0, \frac{2R_\mathrm{e}}{L})$ and an increasing function for $\omega_\mathrm{f} \in ( \frac{2R_\mathrm{e}}{L},\infty)$. There exists an optimal $\omega_\mathrm{f,opt} =2R_\mathrm{e}/L$ that requires the minimum amount of capacitance to maintain stability,
\begin{align} \label{coptimal}
    C>C_{\mathrm{opt}} = C_0(\omega_\mathrm{f,opt} = \frac{2R_\mathrm{e}}{L}) = \frac{L(1+\sqrt{1-K/R_\mathrm{e}})}{2KR_\mathrm{e}}.
\end{align}
Because $C_{\mathrm{opt}} \leq C_{\mathrm{base}}$ where equality holds when $P=0$, with optimal virtual inertia, a smaller $C$ than the baseline can guarantee stability. In other words, the system stability is improved if the DC bus capacitance remains the same.

\textbf{Maximum Inertia (with baseline capacitance)}: If a certain amount of inertia is provided with the baseline capacitance, the maximum inertia that can be provided is calculated by solving $C_0(\omega_\mathrm{f,max}) = L/(KR_\mathrm{e})$, which gives
\begin{align} \label{cmaximum}
    \omega_\mathrm{f,max} = R_\mathrm{e}/{L}.
\end{align}
This is the maximum inertia that can be supplied by a droop-controlled converter without negatively affecting stability.

\textbf{Large Inertia}: Virtual inertia larger than \eqref{cmaximum} will destabilize the grid. If the desired virtual inertia is much larger than \eqref{cmaximum}, the capacitance to guarantee stability is increased significantly, which can be seen as $R_\mathrm{e}/\omega_\mathrm{f}^2$ dominates \eqref{C0},
\begin{align} \label{clarge}
    C>C_{\mathrm{large}} \approx \frac{1}{\omega_\mathrm{f}\sqrt{KR_\mathrm{e}}}. 
\end{align}

\begin{figure}[t]
     \centering
     \begin{subfigure}[b]{0.241\textwidth}
         \centering
         \includegraphics[width=\textwidth]{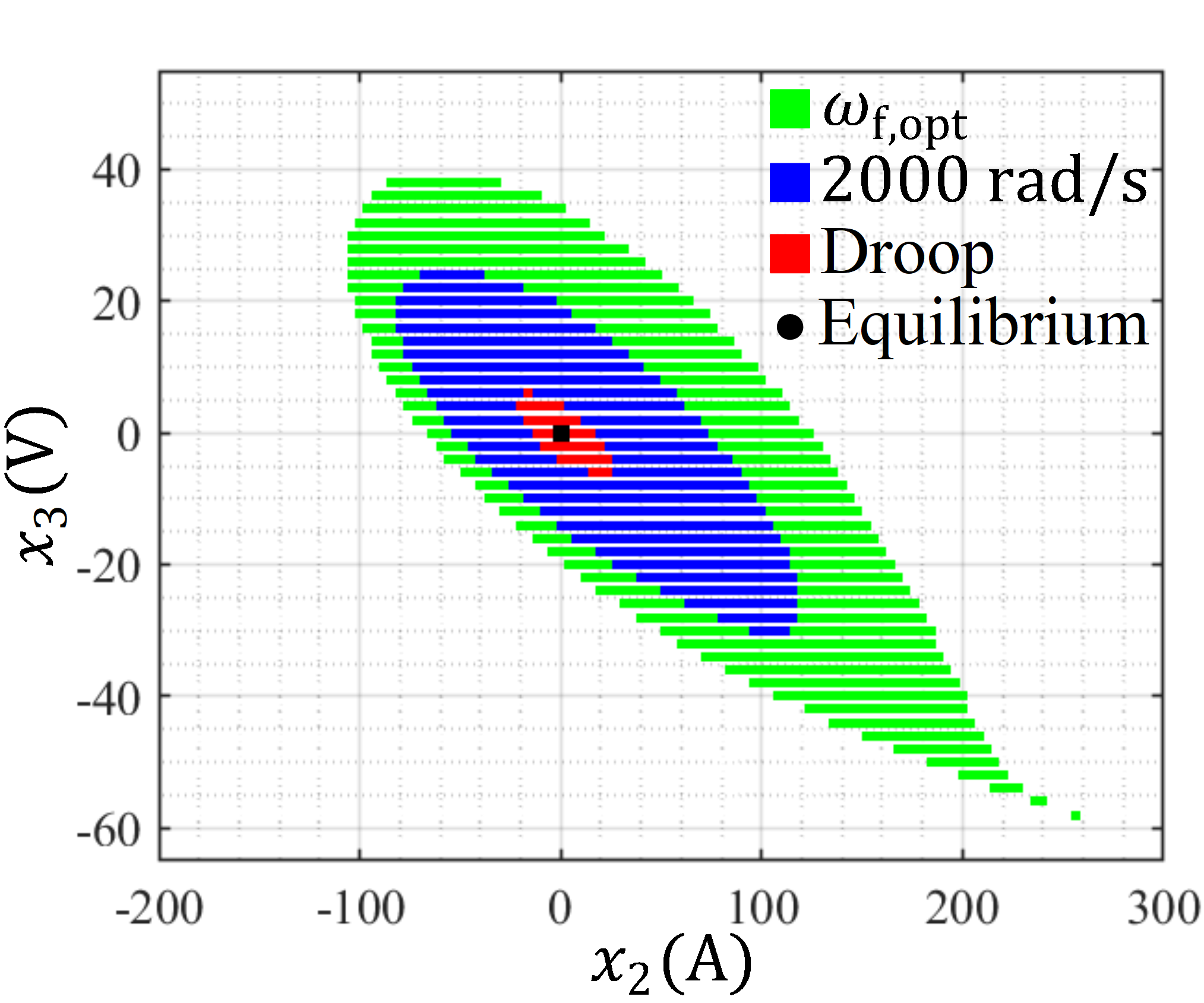}
         \caption{Droop control, $2000$ rad/s, and  $\omega_\mathrm{f,opt}$. }
         \label{fig:roa1}
     \end{subfigure}
     \begin{subfigure}[b]{0.241\textwidth}
         \centering
         \includegraphics[width=\textwidth]{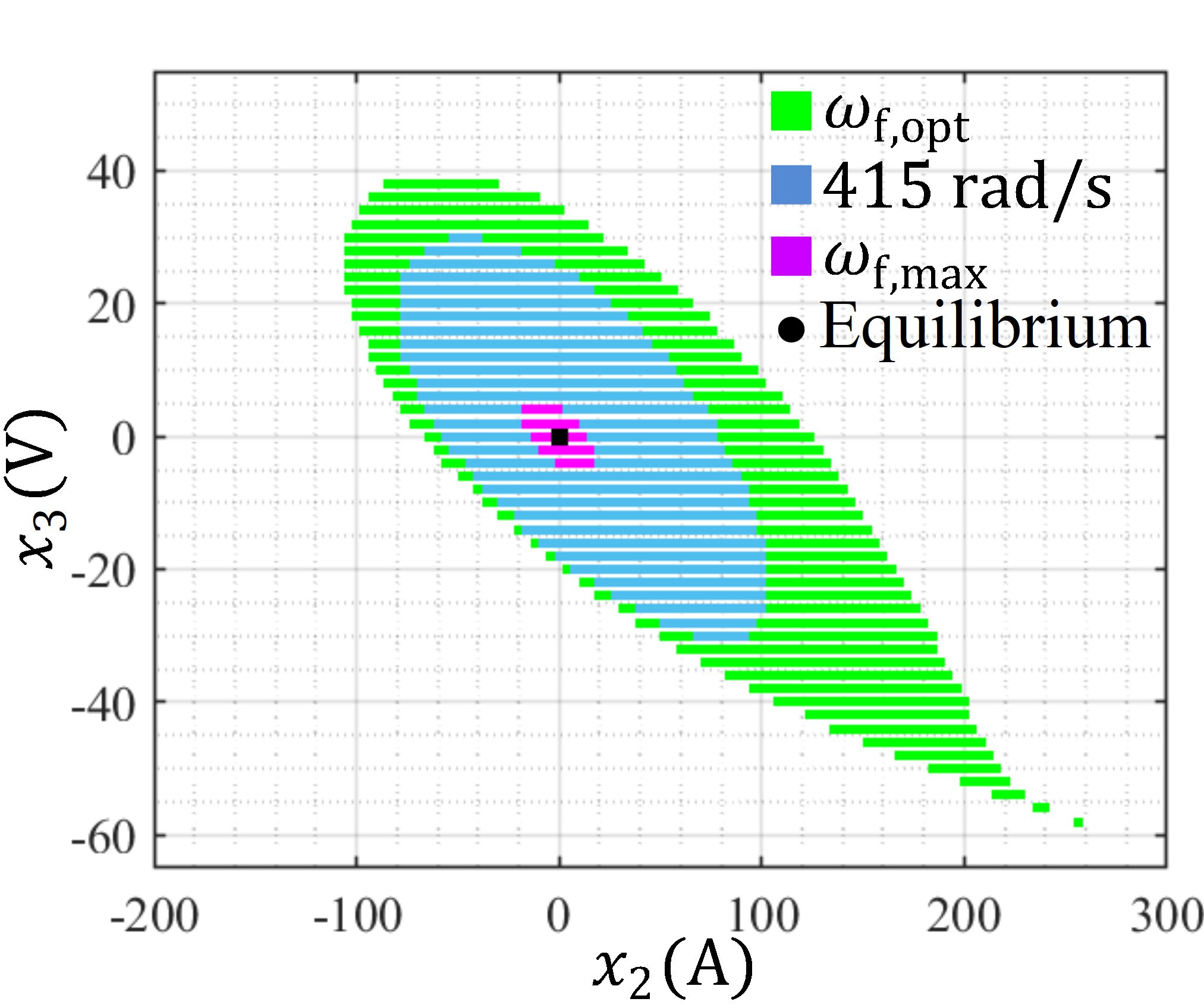}
         \caption{$\omega_\mathrm{f,opt}$, $415$ rad/s, and  $\omega_\mathrm{f,max}$.}
         \label{fig:roa2}
     \end{subfigure}
        \caption{ROAs with different virtual inertia designs.}
        \label{fig:roa}
        \vspace{-15pt}
\end{figure}

\vspace{-10pt}
\subsection{Numerical Verification for Large-Signal Stability}\label{largesignal}
In this subsection, we perform large-signal stability analysis by numerically evaluating the system's ROA with different virtual inertia designs. For convenience, we shift the equilibrium of the system~\eqref{system_model} to the origin by letting $v_\mathrm{ref}=V_\mathrm{ref,e}+x_1$, $i=I_\mathrm{e}+x_2$, and $v=V_\mathrm{e}+x_3$. The system~\eqref{system_model} is equivalent to 
\begin{subequations}\label{large_model}
\begin{align}
    \label{large_model1}
    \dot  x_1 &=-\omega_\mathrm{f}x_1 -\omega_\mathrm{f} K x_2, \\\label{large_model2}
    L\dot x_2 &= x_1 - x_3, \\ \label{large_model3}
    C \dot x_3 & = x_2 + \frac{P}{V_\mathrm{e}(V_\mathrm{e}+x_3)}x_3.
\end{align}
\end{subequations}
Note that \eqref{large_model} is not a small-signal model of \eqref{system_model} because the CPL nonlinearity is captured in~\eqref{large_model3}.

The goal of large-signal stability analysis is to estimate the ROA of the equilibrium point, i.e., $x_1=x_2=x_3=0$. In this paper, we numerically solve \eqref{large_model} with different initial points because this method gives an estimation close to the actual ROA~\cite{6031929}. The drawback is its heavy computational burden since it repeatedly solves the differential equations. %A initial point $(x_1,x_2,x_3)$ is in the ROA if the system converges to the equilibrium point $x_1=x_2=x_3=0$.
The system parameters are given in Table~\ref{Table1}. The estimated ROAs for different virtual inertia designs are shown in Fig.~\ref{fig:roa}. The ROAs are presented in the $x_2-x_3$ plane, which is the most widely used selection for visualization~\cite{6031929,9855660, 7182770} as $x_2$ and $x_3$ represent the current and voltage disturbances, respectively.

For droop control without virtual inertia (i.e., the baseline), the ROA is very small as shown by the red region in Fig.~\ref{fig:roa}(a), which agrees with the results in Fig.~\ref{fig:C0} as $\omega_\mathrm{f}\to\infty$ because the system is close to the stability boundary with $C$ being slightly larger than $C_{\mathrm{base}}$. When virtual inertia is increased, the ROA increases, reaching the largest area around $\omega_\mathrm{f,opt}=715$ rad/s, as shown in  Fig.~\ref{fig:roa}(a). Fig.~\ref{fig:roa}(b) shows that increasing the virtual inertia further reduces the ROA, in agreement with the small-signal analysis summarized in Fig.~\ref{fig:C0}. 

For the system given in Table~\ref{Table1}, numerical results suggest that the largest ROA is achieved by multiple values of $\omega_\mathrm{f}$ around $\omega_\mathrm{f,opt}$~(including $\omega_\mathrm{f,opt}$). However, it remains unknown whether this is true or not for a generic DC grid due to the lack of rigorous large-signal stability analysis.

\vspace{-5pt}
\section{Proposed Design Guideline}\label{design_sec}
\begin{figure}[t]
  \centering
	\includegraphics[width=0.485\textwidth]{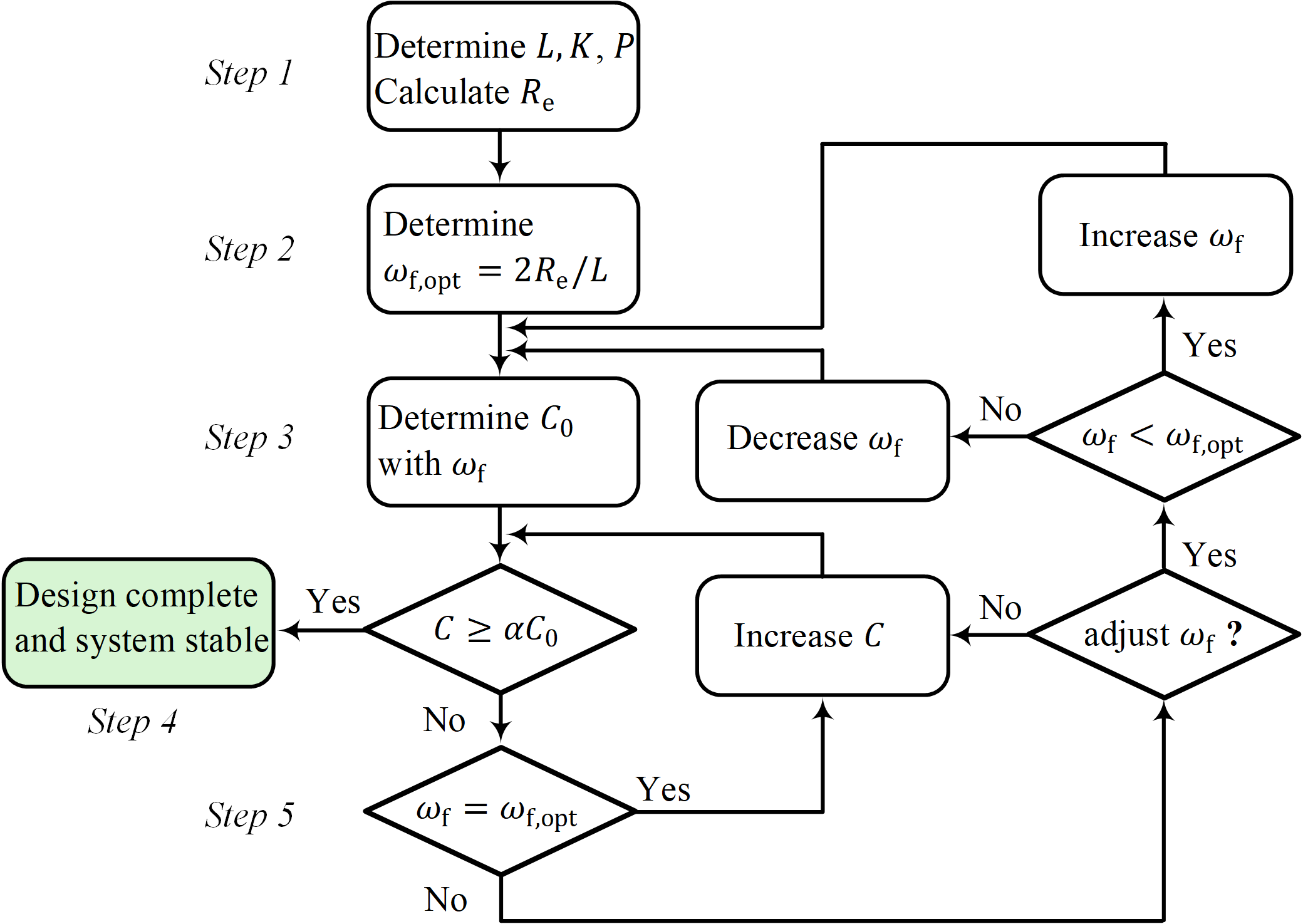}
	\vspace{-10pt}
	\caption{Flowchart to design a stable DC grid with virtual inertia.}
	\label{fig:design_flow}
 \vspace{-5pt}
\end{figure}

A step-by-step guideline based on the proposed stability criterion is provided to design a stable DC Grid with virtual inertia. The guideline is presented as a flowchart in Fig.~\ref{fig:design_flow}.

\textit{Step 1}: Simplify the system model to the form shown in Fig.~\ref{fig:system}, using the method described in~\cite{6909049}; determine parameters $L$, $K$, $P$, and calculate $R_\mathrm{e}$. Inductance $L$ can be calculated based on the geometry of the system conductors, from datasheets or by measurement. Desired maximum voltage deviation determines the droop gain $K$ as discussed in~\cite{6953474}. Power $P$ is the estimated maximum CPL power. It is well-known that stability deteriorates as the CPL load increases~\cite{kwasinski2010dynamic,6909049}. Therefore, the system is designed to be stable for the maximum CPL power with some stability margin, which guarantees stability for any CPL power smaller than the maximum; this will be validated by test results in Section V. 

\textit{Step 2}: Determine the optimal inertia as $\omega_\mathrm{f,opt}=2R_\mathrm{e}/L$.

\textit{Step 3}: Determine the capacitance at the stability boundary $C_0$ based on the desired virtual inertia $\omega_f$, $L$, $K$, and $R_\mathrm{e}$ according to \eqref{C0}.

\textit{Step 4}: If the actual DC bus capacitance $C$ satisfies $C\geq \alpha C_0$, where $\alpha > 1 $ is used to provide a sufficient stability margin, then the design is complete. In this paper, we use $\alpha= 1.3$, which provides a good trade-off between the stability margin and the capacitor size. If a larger stability margin is desired, a larger $\alpha$ can be used.

\textit{Step 5}: If $C< \alpha C_0$, the system will not have the desired stability margin. The desired margin can be restored by either (1) increasing the size of the capacitor $C$, or (2) adjusting the virtual inertia to a value closer to $\omega_\mathrm{f,opt}$ and going back to step 3 to re-calculate $C_0$.

In practical systems, the parasitic resistance, which is neglected in the design guideline, will damp the system to improve stability~\cite{5764841}, providing an additional safety margin.

\begin{figure}[t]
  \centering
  \includegraphics[width=0.36\textwidth]{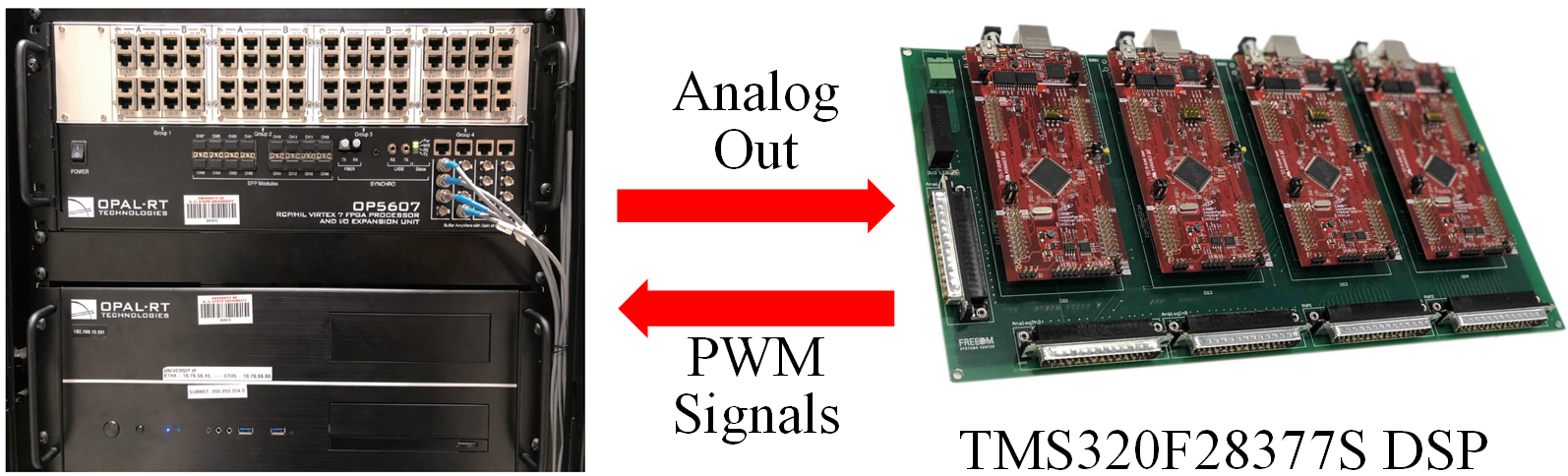}
	\vspace{0pt}
	\caption{HIL testbed.}
	\label{fig:hil}
        \vspace{-10pt}
\end{figure}

\begin{figure}[t]
     \centering
     \begin{subfigure}[b]{0.241\textwidth}
         \centering
         \includegraphics[width=\textwidth]{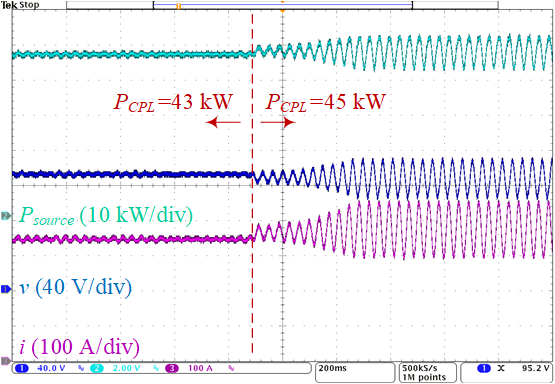}
         \caption{Baseline without virtual inertia.}
         \label{fig:droop}
     \end{subfigure}
     \begin{subfigure}[b]{0.241\textwidth}
         \centering
         \includegraphics[width=\textwidth]{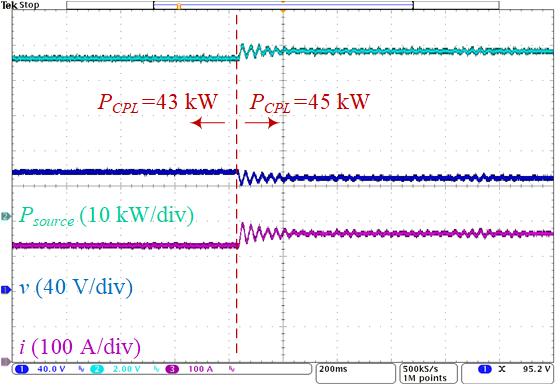}
         \caption{Optimal inertia  $\omega_\mathrm{f,opt} = 715$ rad/s.}
         \label{fig:opt}
     \end{subfigure}
     \begin{subfigure}[b]{0.241\textwidth}
         \centering
         \includegraphics[width=\textwidth]{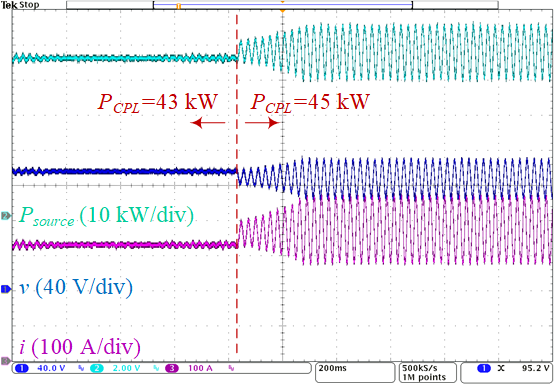}
         \caption{Maximum inertia $\omega_\mathrm{f,max}'= 404$ rad/s.}
         \label{fig:max}
     \end{subfigure}
     \begin{subfigure}[b]{0.241\textwidth}
         \centering
         \includegraphics[width=\textwidth]{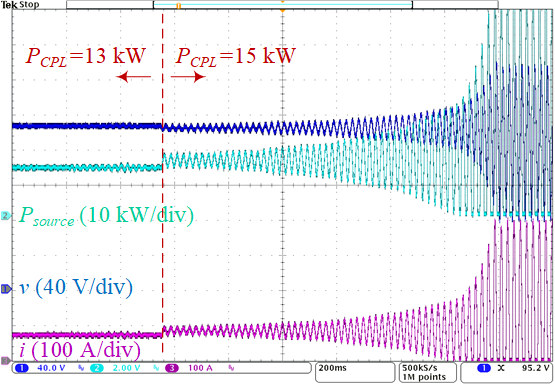}
         \caption{Large inertia $\omega_\mathrm{f,large} = 125$ rad/s.}
         \label{fig:large}
     \end{subfigure}
        \caption{Test results for a single source converter with different virtual inertia.}
        \label{fig:results}
        \vspace{-10pt}
\end{figure}

\section{Experiment Validation}\label{experiment}

\subsection{Analysis Validation with a Single Source Converter}
We validate the analysis using a hardware-in-the-loop (HIL) testbed shown in Fig.~\ref{fig:hil}. An OPAL-RT real-time simulator is used to model the system in Fig.~\ref{fig:system}. The system parameters are given in Table~\ref{Table1}. The source converter is a buck converter with inner current and voltage loops to regulate its output voltage. The CPL is a point-of-load buck converter regulating its power output. Digital signal processors TMS320F28377S from Texas Instruments are used to control the source converter and the CPL. The test results are shown in Fig.~\ref{fig:results}.

First, we evaluate the baseline, i.e., droop control without virtual inertia. With the parameters given in Table~\ref{Table1}, the baseline capacitance is $C_{\mathrm{base}} \approx 14$~mF according to \eqref{cbaseline}. With $C=14$~mF, the proposed stability criterion predicts that system will be unstable if the CPL power is larger than $46$~kW. In the test, we increase the CPL power with $2$~kW steps. Fig.~\ref{fig:results}(a) shows the test results when the CPL power is stepped from $43$~kW to $45$~kW. The system becomes oscillatory, which validates the instability caused by the CPL and our analysis. The actual CPL power $45$~kW causing instability is smaller than the predicted $46$~kW mainly because the equilibrium near the instability boundary has a small ROA, and perturbations in practical operation can make the system unstable.

We add virtual inertia in the source converter control and set the LPF bandwidth to the optimal inertia $\omega_\mathrm{f,opt} = 715$~rad/s. In this case, the minimum capacitance to guarantee stability is $C_{\mathrm{opt}}=11.6$~mF according to \eqref{coptimal}. The system stability is improved since $C>C_{\mathrm{opt}}$. Fig.~\ref{fig:results}(b) shows the results for the same test. The transient oscillation is damped, and the system is stable at $P_{\mathrm{CPL}}=45$~kW.

Next, we reduce the LPF bandwidth to $\omega_\mathrm{f,max}'= 404$~rad/s to provide larger inertia. Fig.~\ref{fig:results}(c) shows the test results, which present similar oscillatory behavior at $P_{\mathrm{CPL}}=45$~kW as the baseline results. It is worth noting that inertia with $\omega_\mathrm{f,max}'$ is smaller than maximum inertia predicted by \eqref{cmaximum} with $\omega_\mathrm{f,max}= 357$~rad/s. This is mainly because the source converter is not an ideal voltage source and has dynamics from the inner loops. 

Finally, we further reduce the LPF bandwidth to  $\omega_\mathrm{f,large} = 125$~rad/s to provide even larger virtual inertia, and the test results are shown in Fig.~\ref{fig:results}(d). The system becomes unstable when the CPL power is stepped from $13$~kW to $15$~kW, showing that the system stability is significantly deteriorated. 

\subsection{Design Guideline Validation}

\begin{figure}[t]
     \centering
     \begin{subfigure}[b]{0.241\textwidth}
         \centering
         \includegraphics[width=\textwidth]{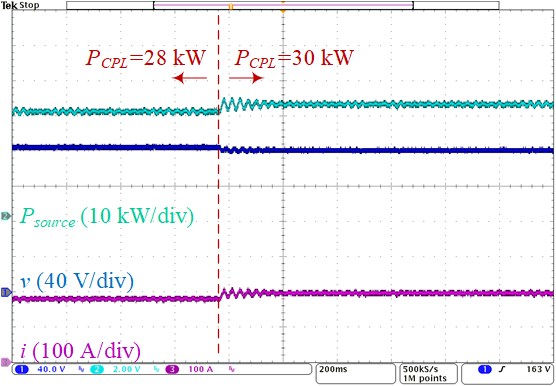}
         \caption{CPL power steps to $30$~kW.}
         \label{design1}
     \end{subfigure}
     \begin{subfigure}[b]{0.241\textwidth}
         \centering
         \includegraphics[width=\textwidth]{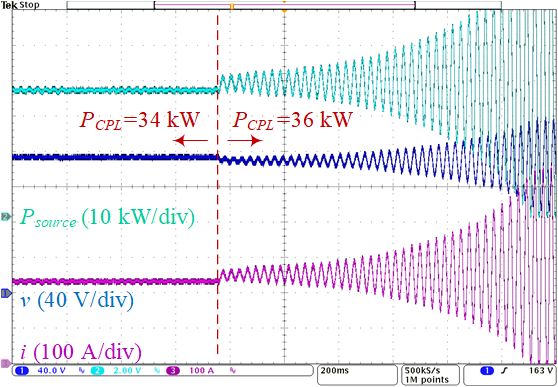}
         \caption{CPL power steps to $36$~kW.}
         \label{design2}
     \end{subfigure}
        \caption{Test results with $C=27$ mF and $\omega_\mathrm{f}=125$ rad/s.}
        \label{design}
        \vspace{-13pt}
\end{figure}

To validate the design guidelines, we consider a system with $K$, $L$ amd $P$ given in Table~\ref{Table_design}, and the desired virtual inertia $\omega_\mathrm{f}=125$~rad/s. Following the guideline, step 2 determines the optimal inertia as $\omega_\mathrm{f,opt} = 1777$~rad/s. In step 3, $C_0= 20.6$~mF is calculated with $\omega_\mathrm{f}=125$~rad/s. The condition in step 4 is not satisfied as initially $C=14$~mF $<\alpha C_0 = 26.8$~mF, where $\alpha=1.3$ is used to provide a stability margin. According to step 5, one way to improve stability is to increase $\omega_\mathrm{f}$. However, in this example, the desired virtual inertia is fixed at $\omega_\mathrm{f}=125$~rad/s. Therefore, we increase the capacitance to $C=27$~mF such that the condition~$C\geq \alpha C_0$ is satisfied. The test results with $C=27$~mF are shown in Fig.~\ref{design}.

\begin{table}[t]
\footnotesize
\centering\captionsetup{justification=centering}
\caption{System parameters for validating the design guideline}\label{Table_design}
\begin{tabular}{cccccc}%{|p{2.5cm}|p{2.5cm}|p{2.5cm}|}
\hline
\hline
$V_\mathrm{n} (\mathrm{V})$ & $K$ & $L (\mathrm{mH})$ & $C (\mathrm{mF})$ & $P (\mathrm{kW})$ & $\omega_\mathrm{f}$ (rad/s)\\
\hline
\multirow{2}{*}{$200$} &\multirow{2}{*}{$0.2$} & \multirow{2}{*}{$1$} & $14$ (Before) & \multirow{2}{*}{$30$} &  \multirow{2}{*}{$125$} \\
& & &$27$ (Designed) & &\\
\hline
\hline
\end{tabular}
\vspace{-5pt}
\end{table}

To test out the design guideline, we increase the CPL power with 2 kW steps. Fig.~\ref{design}(a) shows the test results when the CPL power is stepped from $28$~kW to $30$~kW. The transient oscillation is damped, and the system is stable, thus meeting the design objective of supplying the CPL up to $30$~kW while providing virtual inertia with $\omega_\mathrm{f}=125$~rad/s. If we further increase the CPL power, the system becomes unstable when CPL power reaches $36$~kW as shown in Fig.~\ref{design}(b). Therefore, the designed $30$~kW system has a $6$~kW stability margin.

\begin{table}[t]
\footnotesize
\centering\captionsetup{justification=centering}
\caption{System parameters for the multiple-converter case}\label{Table2}
\begin{tabular}{cccccc}%{|p{2.5cm}|p{2.5cm}|p{2.5cm}|}
\hline
\hline
&$V_\mathrm{n}~(\mathrm{V})$ & $K$ & $L (\mathrm{mH})$ & $C~(\mathrm{mF})$ & $P (\mathrm{kW})$\\
\hline
Source 1&$200$ &$1$ & $4$ &  &  \\
Source 2&$200$ &$1$ & $4$ &  &  \\
Source 3&$200$ &$0.5$ & $2$ &  &  \\
\hline
DC bus and CPL & & &  & 8 & 36 \\
\hline
\hline
\end{tabular}
\end{table}

\begin{figure}[t]
  \centering
	\includegraphics[width=0.44\textwidth]{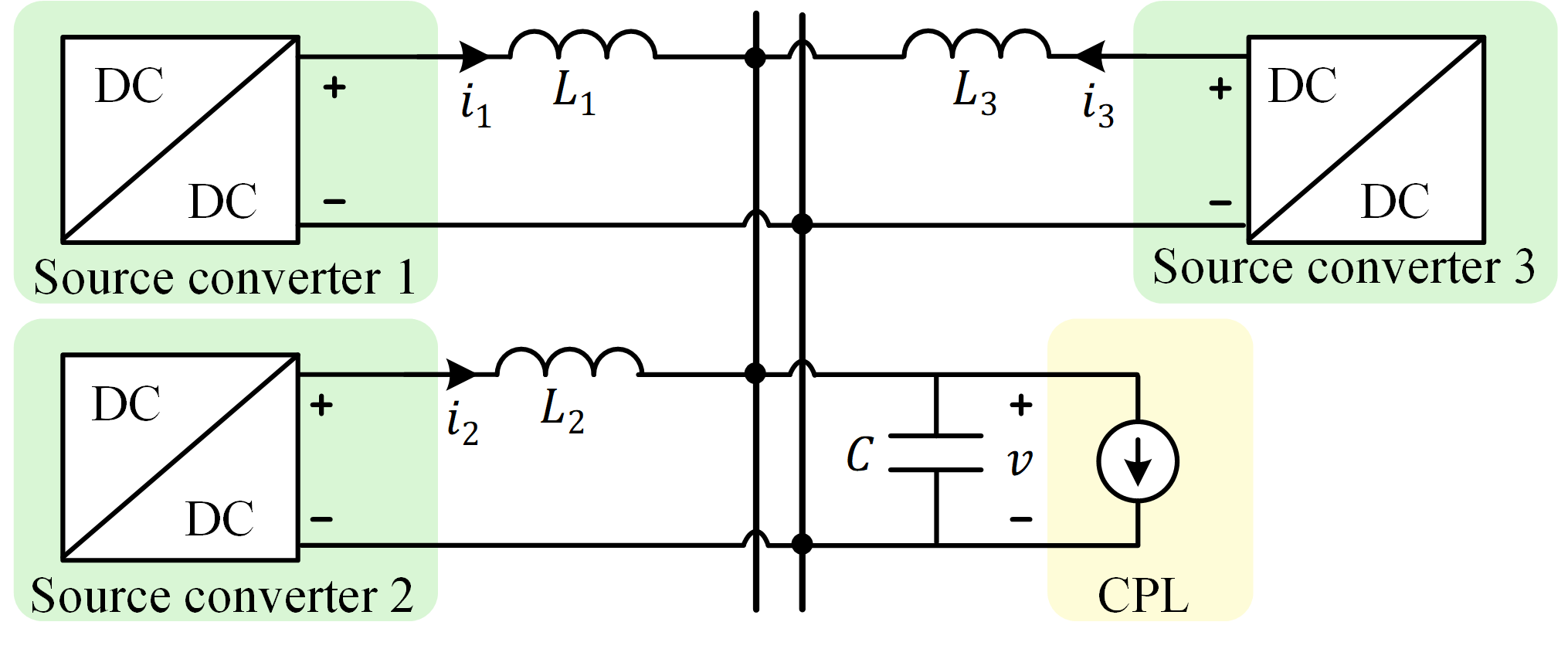}
	\caption{DC grid under test with three source converters.}
	\label{fig:system3}
	\vspace{+12pt}
     \centering
     \begin{subfigure}[b]{0.241\textwidth}
         \centering
         \includegraphics[width=\textwidth]{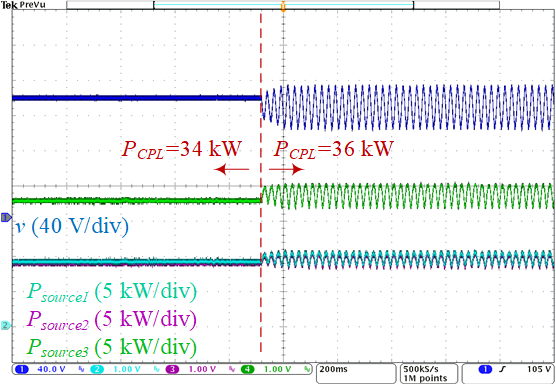}
         \caption{Baseline without virtual inertia.}
         \label{fig:droop3}
     \end{subfigure}
     \begin{subfigure}[b]{0.241\textwidth}
         \centering
         \includegraphics[width=\textwidth]{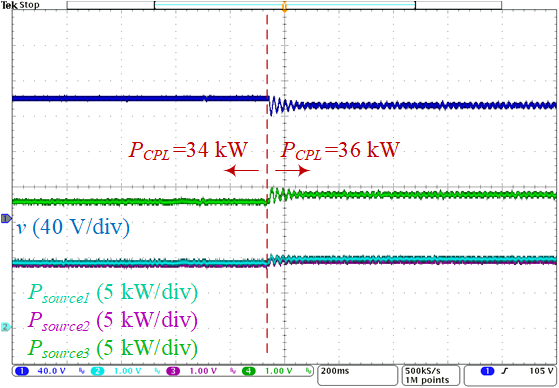}
         \caption{Optimal inertia  $\omega_\mathrm{f,opt} = 1049$~rad/s.}
         \label{fig:opt3}
     \end{subfigure}
     \begin{subfigure}[b]{0.241\textwidth}
         \centering
         \includegraphics[width=\textwidth]{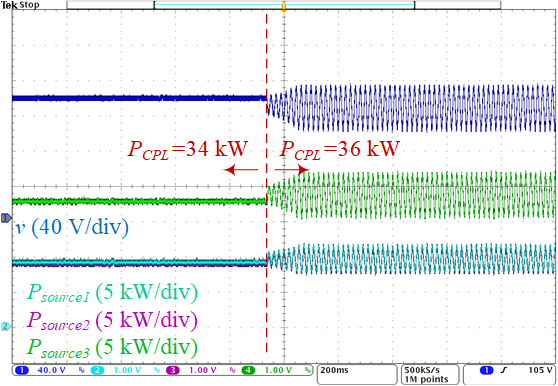}
         \caption{Maximum inertia $\omega_\mathrm{f,max}'= 591$~rad/s.}
         \label{fig:max3}
     \end{subfigure}
     \begin{subfigure}[b]{0.241\textwidth}
         \centering
         \includegraphics[width=\textwidth]{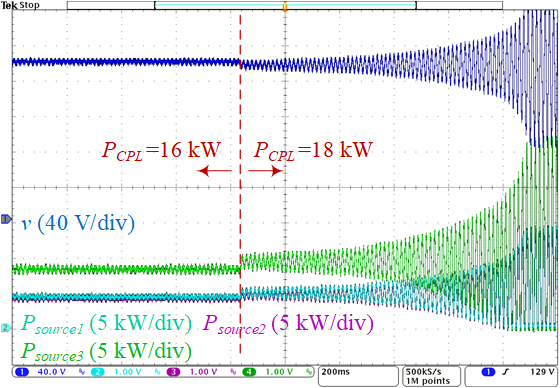}
         \caption{Large inertia $\omega_\mathrm{f,large} = 251$~rad/s.}
         \label{fig:large3}
     \end{subfigure}
        \caption{Test results for three source converters with different virtual inertia.}
        \label{fig:results3}
        \vspace{-10pt}
\end{figure}

\subsection{Analysis Validation with Multiple Source Converters}
We consider a DC grid with three source converters feeding a CPL, as shown in Fig.~\ref{fig:system3}. The system parameters are given in Table~\ref{Table2}. The source converters are designed to have a $1:1:2$ power sharing. The test results are shown in Fig.~\ref{fig:results3}.

Using the method in~\cite{6909049}, a single-converter equivalent model can be built for the three source converters. According to \cite{6909049}, the equivalent inductance and equivalent droop gain are calculated by
\begin{align}
    L_\mathrm{eq}=\frac{1}{
    \sum_{i=1}^3\frac{1}{L_i}}=1~\mathrm{mH}, \
    K_\mathrm{eq}=\frac{L_\mathrm{eq}}{\sum_{i=1}^3L_i}\sum_{i=1}^3 K_i=0.25.
\end{align}

With $K_\mathrm{eq}$, $L_\mathrm{eq}$, and $P=36$~kW, the baseline capacitance is $C_{\mathrm{base}} \approx 8$~mF according to \eqref{cbaseline}. With $C=8$~mF, the grid will be unstable if the CPL power exceeds $36$~kW. Fig.~\ref{fig:results3}(a) shows the results for baseline without virtual inertia. When $P_{\mathrm{CPL}}=34$~kW, the grid is stable and the power sharing of the three source converters is 1:1:2 as designed. When the CPL power is stepped to $36$~kW, the grid becomes unstable.

Fig.~\ref{fig:results3}(b) shows the results when the optimal inertia $\omega_\mathrm{f,opt} = 1047$~rad/s is implemented in all source converters. The oscillation is damped, and the system is stable at $P_{\mathrm{CPL}}=36$~kW. In the steady state, the power sharing is 1:1:2.

We reduce the LPF bandwidth to $\omega_\mathrm{f,max}'= 591$~rad/s to provide larger inertia. Fig.~\ref{fig:results3}(c) shows the test results. Instability occurs at $P_{\mathrm{CPL}}=36$~kW, which is the same as the baseline results in Fig.~\ref{fig:results3}(a). Similar to the single-converter case, the inertia with $\omega_\mathrm{f,max}'$ is smaller than the value predicted by \eqref{cmaximum} ($\omega_\mathrm{f,max} = 523$~rad/s) mainly because the source converters are not ideal voltage sources. 

Finally, we further reduce the LPF bandwidth to $\omega_\mathrm{f,large} = 251$~rad/s to provide even larger virtual inertia. The test results in Fig.~\ref{fig:results3}(d) show the system becomes unstable when the CPL power is stepped from $16$~kW to $18$~kW. 

The above test results show that the presented analysis applies to DC grids with multiple source converters with the same virtual inertia parameter $\omega_\mathrm{f}$. It is worth mentioning that setting the same $\omega_\mathrm{f}$ for all source converters in a DC grid helps maintain proportional power sharing during transients and is a common design practice in the literature~(e.g.~\cite{9264685}). 

\section{Conclusion}\label{conclusion}
Virtual inertia has a significant impact on DC grid stability with CPLs. In this paper, we derive a generalized stability criterion for converters operating with virtual inertia and driving a CPL. We show that the system stability is a function of the droop gain $K$, negative load equivalent impedance $R_\mathrm{e}$, virtual inertia $\omega_\mathrm{f}$, and the system parameters $L$ and $C$. We derive a closed-form expression that links the DC bus capacitance to the virtual inertia parameter. The analysis shows that, for a fixed system design, there is a range of virtual inertia values that improve stability. If the inertia is further increased, beyond a maximum value, the system becomes unstable. 

We validate the analysis using a HIL testbed for a single source converter connected to a CPL and for a DC grid with three source converters driving a CPL. In both cases, the results match closely with the predicted performance.

\vspace{-0pt}

% use section* for acknowledgment

\iffalse
\section*{Acknowledgment}
The authors would like to thank...
\fi

% Can use something like this to put references on a page
% by themselves when using endfloat and the captionsoff option.
\ifCLASSOPTIONcaptionsoff
  \newpage
\fi

% trigger a \newpage just before the given reference
% number - used to balance the columns on the last page
% adjust value as needed - may need to be readjusted if
% the document is modified later
%\IEEEtriggeratref{8}
% The "triggered" command can be changed if desired:
%\IEEEtriggercmd{\enlargethispage{-5in}}

% references section

% can use a bibliography generated by BibTeX as a .bbl file
% BibTeX documentation can be easily obtained at:
% http://mirror.ctan.org/biblio/bibtex/contrib/doc/
% The IEEEtran BibTeX style support page is at:
% http://www.michaelshell.org/tex/ieeetran/bibtex/
\bibliographystyle{IEEEtran}
% argument is your BibTeX string definitions and bibliography database(s)
\bibliography{reference}
\end{document}